\documentclass{svmult}

\usepackage[english]{babel}
\usepackage[utf8]{inputenc}
\usepackage{amsmath}       
\usepackage{amssymb}
\usepackage{graphicx}

\usepackage{enumitem}
\usepackage{multirow}

\usepackage{makeidx}         
\usepackage{graphicx}        
\usepackage{multicol}        
\usepackage[bottom]{footmisc}
\usepackage{url}

\makeindex  

\begin{document}
\title*{``Market making'' in an order book model and its impact on the spread}
\author{Ioane Muni Toke}
\institute{Chair of Quantitative Finance, Applied Mathematics and Systems Laboratory, Ecole Centrale Paris, Grande Voie des Vignes, 92290 Chatenay-Malabry, France. \texttt{ioane.muni-toke@ecp.fr}}
\date{\today}

\maketitle

\begin{abstract}
It has been suggested that marked point processes might be good candidates for the modelling of financial high-frequency data. A special class of point processes, Hawkes processes, has been the subject of various investigations in the financial community. In this paper, we propose to enhance a basic zero-intelligence order book simulator with arrival times of limit and market orders following mutually (asymmetrically) exciting Hawkes processes. Modelling is based on empirical observations on time intervals between orders that we verify on several markets (equity, bond futures, index futures). We show that this simple feature enables a much more realistic treatment of the bid-ask spread of the simulated order book.
\end{abstract}

\section*{Introduction}

\paragraph{Arrival times of orders: event time is not enough}

As of today, the study of arrival times of orders in an order book has not been a primary focus in order book modelling. Many toy models leave this dimension aside when trying to understand the complex dynamics of an order book. In most order driven market models such as \cite{ContBouchaud2000, LuxMarchesi2000, Pietronero2009}, and in some order book models as well (e.g.\cite{Preis2006}), a time step in the model is an arbitrary unit of time during which many events may happen. We may call that clock \textit{aggregated time}.
In most order book models such as \cite{ChalletStinchcombe2001, MikeFarmer2008, ContStoikov2009}, one order is simulated per time step with given probabilities, i.e. these models use the clock known as \textit{event time}. In the simple case where these probabilities are constant and independent of the state of the model, such a time treatment is equivalent to the assumption that order flows are homogeneous Poisson processes.
A probable reason for the use of non-physical time in order book modelling -- leaving aside the fact that models can be sufficiently complicated without adding another dimension -- is that many puzzling empirical observations (see e.g. \cite{ChakrabortiMuniToke2009} for a review of some of the well-known ``stylized facts'') can (also) be done in event time (e.g. autocorrelation of the signs of limit and market orders) or in aggregated time (e.g. volatility clustering). 

However, it is clear that physical (calendar) time has to be taken into account for the modelling of a realistic order book model. For example, market activity varies widely, and intraday seasonality is often observed as a well known U-shaped pattern. Even for a short time scale model -- a few minutes, a few hours -- durations of orders (i.e. time intervals between orders) are very broadly distributed. Hence, the Poisson assumption and its exponential distribution of arrival times has to be discarded, and models must take into account the way these irregular flows of orders affect the empirical properties studied on order books. 

\begin{figure}[ht]
\begin{center}
\rotatebox{270}{
\includegraphics[width=0.5\textwidth]{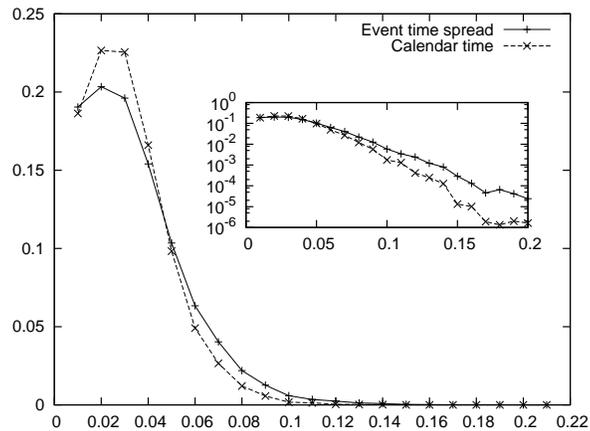}	
}
\end{center}
\caption{Empirical density function of the distribution of the bid-ask spread in event time and in physical (calendar) time. In inset, same data using a semi-log scale. This graph has been computed with 15 four-hour samples of tick data on the BNPP.PA stock (see section \ref{subsection:Data} for details).}
\label{figure:SpreadEventWeighted}
\end{figure}

Let us give one illustration. On figure~\ref{figure:SpreadEventWeighted}, we plot examples of the empirical density function of the observed spread in event time (i.e. spread is measured each time an event happens in the order book), and in physical (calendar) time (i.e. measures are weighted by the time interval during which the order book is idle).
It appears that density of the most probable values of the time-weighted distribution is higher than in the event time case. Symmetrically, the density of the least probable event is even smaller when physical time is taken into account. This tells us a few things about the dynamics of the order book, which could be summarized as follows: the wider the spread, the faster its tightening.
We can get another insight of this empirical property by measuring on our data the average waiting time before the next event, conditionally on the spread size. When computed on the lower one-third-quantile (small spread), the average waiting time is 320 milliseconds. When computed on the upper one-third-quantile (large spread), this average waiting time is 200 milliseconds. These observations complement some of the ones that can be found in the early paper \cite{Biais1995}.

\paragraph{Counting processes with dynamic intensity}

There is a trend in the econometrics literature advocating for the use of (marked) point processes for the modelling of financial time series. One may find a possible source of this interest in \cite{EngleRussell1997, Engle2000}, which introduce autoregressive conditional duration (ACD) and autoregressive conditional intensity (ACI) models. \cite{HallHautsch2007} fit that type of models on the arrival times of limit, market and cancellation orders in an Australian stock market order book.

A particular class of point processes, known as the Hawkes processes, is of special interest for us, because of its simplicity of parametrization. A univariate linear self-exciting Hawkes process $(N_t)_{t>0}$, as introduced by \cite{Hawkes1971, HawkesOakes1974}, is a point process with intensity:
\begin{equation}
	\lambda(t) = \lambda_0 + \int_0^t \nu(t-s) dN_s,
\end{equation}
where the kernel $\nu$ is usually parametrized as $\nu(t) = \alpha e^{-\beta t}$ or in a more general way $\nu(t) = \sum_{k=1}^{p} \alpha_k t^k e^{-\beta t}$. Statistics of this process are fairly well-known and results for a maximum likelihood estimation can be found in \cite{Ogata1981}. In a multivariate setting, mutual excitation is introduced. A bivariate model can thus be written:
\begin{equation}
\left\{
\begin{array}{rcccc}
	\lambda^1(t) & = & \lambda^1_0 & + \int_0^t \nu_{11}(t-s) dN^1_s & + \int_0^t \nu_{12}(t-s) dN^2_s
	\\
	\lambda^2(t) & = & \lambda^2_0 & + \int_0^t \nu_{21}(t-s) dN^1_s & + \int_0^t \nu_{22}(t-s) dN^2_s
\end{array}
\right.
\end{equation}

The use of these processes in financial modelling is growing. We refer the reader to \cite{BauwensHautsch2009} for a review and \cite{Hautsch2004} for a textbook treatment. In \cite{Bowsher2007}, a bivariate (generalized) Hawkes process is fitted to the time series of trades and mid-quotes events, using trading data of the General Motors stock traded on the New York stock Exchange. In \cite{Large2007} a ten-variate Hawkes process is fitted to the Barclay's order book on the London Stock Exchange, sorting orders according to their type and aggressiveness. It is found that the largest measured effect is the exciting effect of market orders on markets orders. \cite{Hewlett2006} fits a bivariate Hawkes model to the time series of buy and sell trades on the EUR/PLN (Euro/Polish Zlotych) FX market. Using the simplest parametrization of Hawkes processes and some (very) constraining assumptions, some analytical results of trade impact may be derived. \cite{Bacry2010} fits a bivariate Hawkes process to the trade time series of two different but highly correlated markets, the ``Bund'' and the ``Bobl'' (Eurex futures on mid- and long-term interest rates).

\paragraph{Organization of this paper}
In this paper, we propose to enhance a basic order book simulator with arrival times of limit and market orders following mutually (asymmetrically) exciting Hawkes processes. Modelling is based on empirical observations, verified on several markets (equities, futures on index, futures on bonds), and detailed in section~\ref{section:EmpiricalEvidence}. More specifically, we observe evidence of some sort of ``market making'' in the studied order books: after a market order, a limit order is likely to be submitted more quickly than it would have been without the market order. In other words, there is a clear reaction that seems to happen: once liquidity has been removed from the order book, a limit order is triggered to replace it. We also show that the reciprocal effect is not observed on the studied markets. These features lead to the use of unsymmetrical Hawkes processes for the design of an agent-based order book simulator described in section~\ref{section:OrderBookSimulator}. We show in section~\ref{section:NumericalResults} that this simple feature enables a much more realistic treatment of the bid-ask spread of the simulated order book.

\section{Empirical evidence of ``market making''}
\label{section:EmpiricalEvidence}

\subsection{Data and observation setup}
\label{subsection:Data}
We use order book data for several types of financial assets:
\begin{itemize} 
	\item BNP Paribas (RIC\footnote{Reuters Identification Code}: BNPP.PA): 7th component of the CAC40 during the studied period
	\item Peugeot (RIC: PEUP.PA): 38th component of the CAC40 during the studied period
	\item Lagardère SCA (RIC: LAGA.PA): 33th component of the CAC40 during the studied period
	\item Dec.2009 futures on the 3-month Euribor (RIC: FEIZ9)
	\item Dec.2009 futures on the Footsie index (RIC: FFIZ9) 
\end{itemize}

We use Reuters RDTH tick-by-tick data from September 10th, 2009 to September 30th, 2009 (i.e. 15 days of trading). For each trading day, we use only 4 hours of data, precisely from 9:30 am to 1:30 pm. As we are studying European markets, this time frame is convenient because it avoids the opening of American markets and the consequent increase of activity.

Our data is composed of snapshots of the first five limits of the order books (ten for the BNPP.PA stock). These snapshots are timestamped to the millisecond and taken at each change of any of the limits or at each transaction. The data analysis is performed as follows for a given snapshot: 
\begin{enumerate}
	\item if the transaction fields are not empty, then we record a market order, with given price and volume;
	\item if the quantity offered at a given price has increased, then we record a limit order at that price, with a volume equal to the difference of the quantities observed;
	\item if the quantity offered at a given price has decreased without any transaction being recorded, then we record a cancellation order at that price, with a volume equal to the difference of the quantities observed;
	\item finally, if two orders of the same type are recorded at the same time stamp, we record only one order with a volume equal to the sum of the two measured volumes.
\end{enumerate}

Therefore, market orders are well observed since transactions are explicitly recorded, but it is important to note that our measure of the limit orders and cancellation orders is not direct. In table~\ref{table:nOrders}, we give for each studied order book the number of market and limit orders detected on our 15 4-hour samples.
\begin{table}[ht]
\begin{center}
\begin{tabular}{|r|c|c|}
	\hline
	Code & Number of limit orders & Number of market orders\\
	\hline
	BNPP.PA & 321,412 & 48,171 \\
	\hline
	PEUP.PA & 228,422 & 23,888 \\
	\hline
	LAGA.PA & 196,539 & 9,834 \\
	\hline
	FEIZ9 & 110,300 & 10,401 \\
	\hline
	FFIZ9 & 799,858 & 51,020 \\
	\hline
\end{tabular}
\end{center}
\caption{Number of limit and markets orders recorded on 15 samples of four hours (Sep 10th to Sep 30th, 2009 ; 9:30am to 1:30pm) for 5 different assets (stocks, index futures, bond futures)}
\label{table:nOrders}
\end{table}
On the studied period, market activity ranges from 2.7 trades per minute on the least liquid stock (LAGA.PA) to 14.2 trades per minute on the most traded asset (Footsie futures).

\subsection{Empirical evidence of ``market making''}
\label{subsection:EmpiricalEvidence}
Our idea for an enhanced model of order streams is based on the following observation: once a market order has been placed, the next limit order is likely to take place faster than usual. To illustrate this, we compute for all our studied assets:
\begin{itemize}
	\item the empirical probability density function (pdf) of the time intervals of the counting process of all orders (limit orders and market orders mixed), i.e. the time step between any order book event (other than cancellation)
	\item and the empirical density function of the time intervals between a market order and the immediately following limit order.
\end{itemize}
If an independent Poisson assumption held, then these empirical distributions should be identical. However, we observe a very high peak for short time intervals in the second case. The first moment of these empirical distributions is significant: one the  studied assets, we find that the average time between a market order and the following limit order is 1.3 (BNPP.PA) to 2.6 (LAGA.PA) times shorter than the average time between two random consecutive events.

\begin{figure}[ht]
\begin{center}
\begin{tabular}{cc}
	\hspace{-0.5cm}
	\rotatebox{270}{\includegraphics[width=0.365\textwidth]{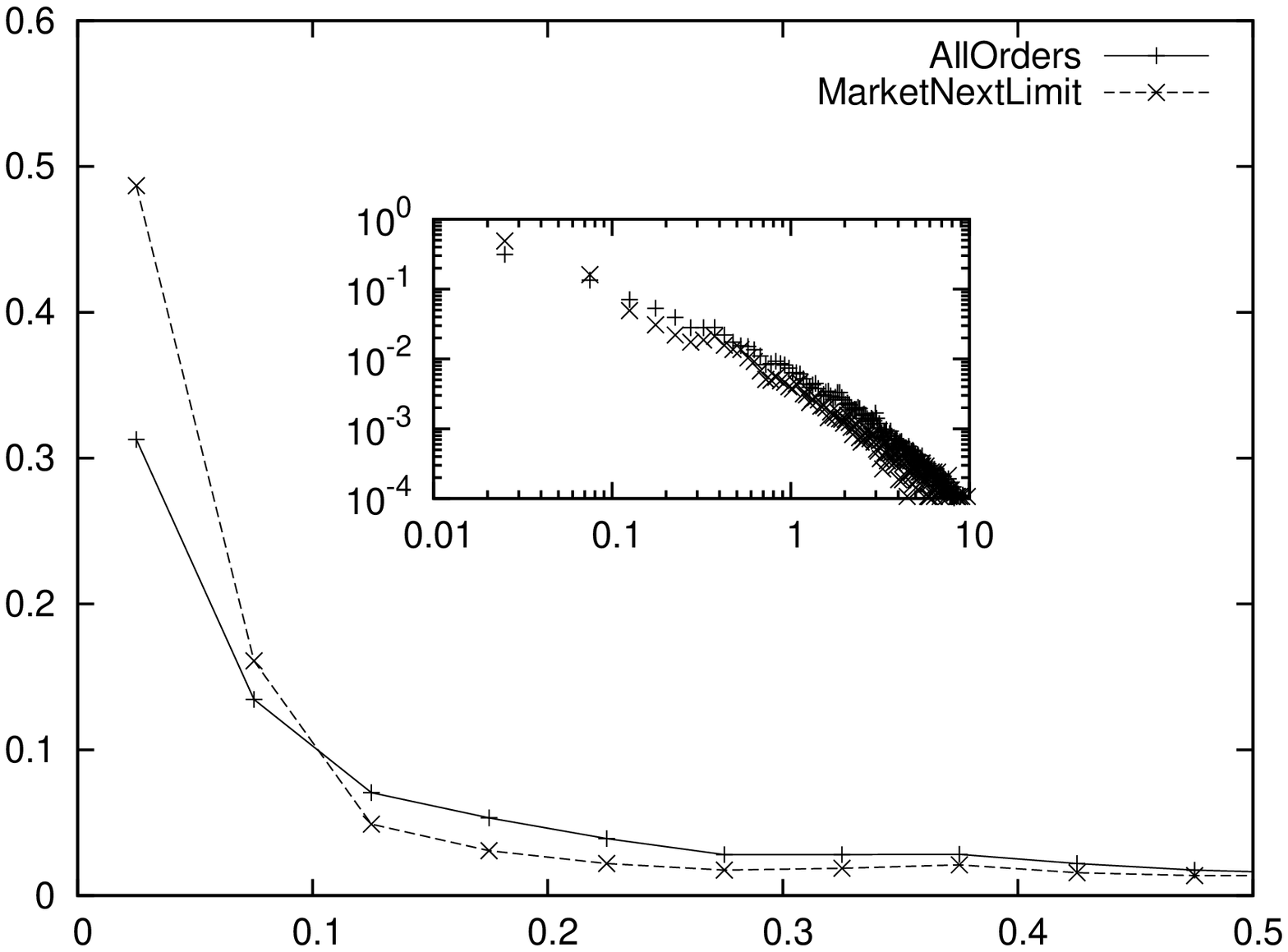}}
	&
	\hspace{-0.5cm}
	\rotatebox{270}{\includegraphics[width=0.365\textwidth]{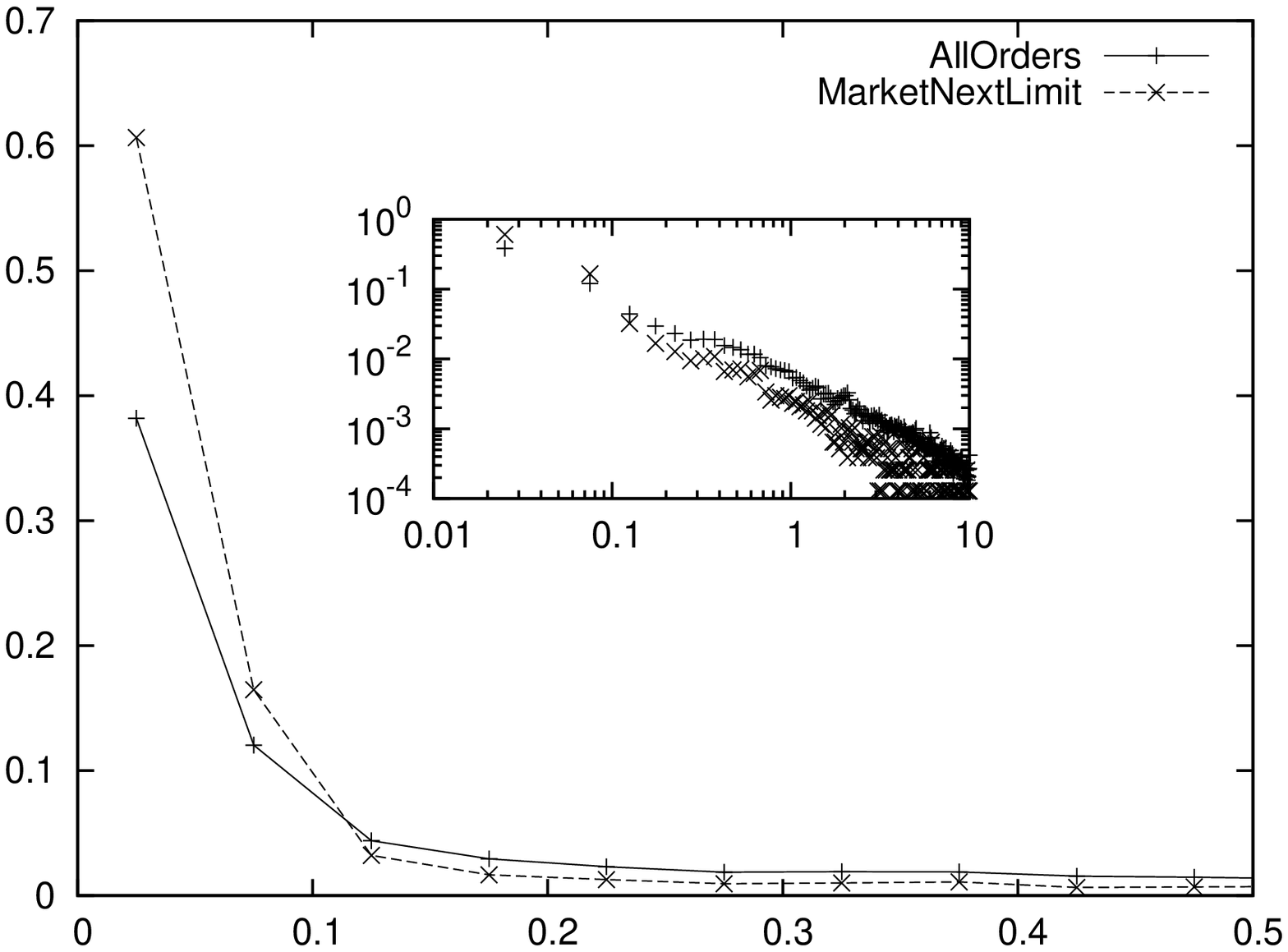}}
	\\
	\hspace{-0.5cm}
	\rotatebox{270}{\includegraphics[width=0.365\textwidth]{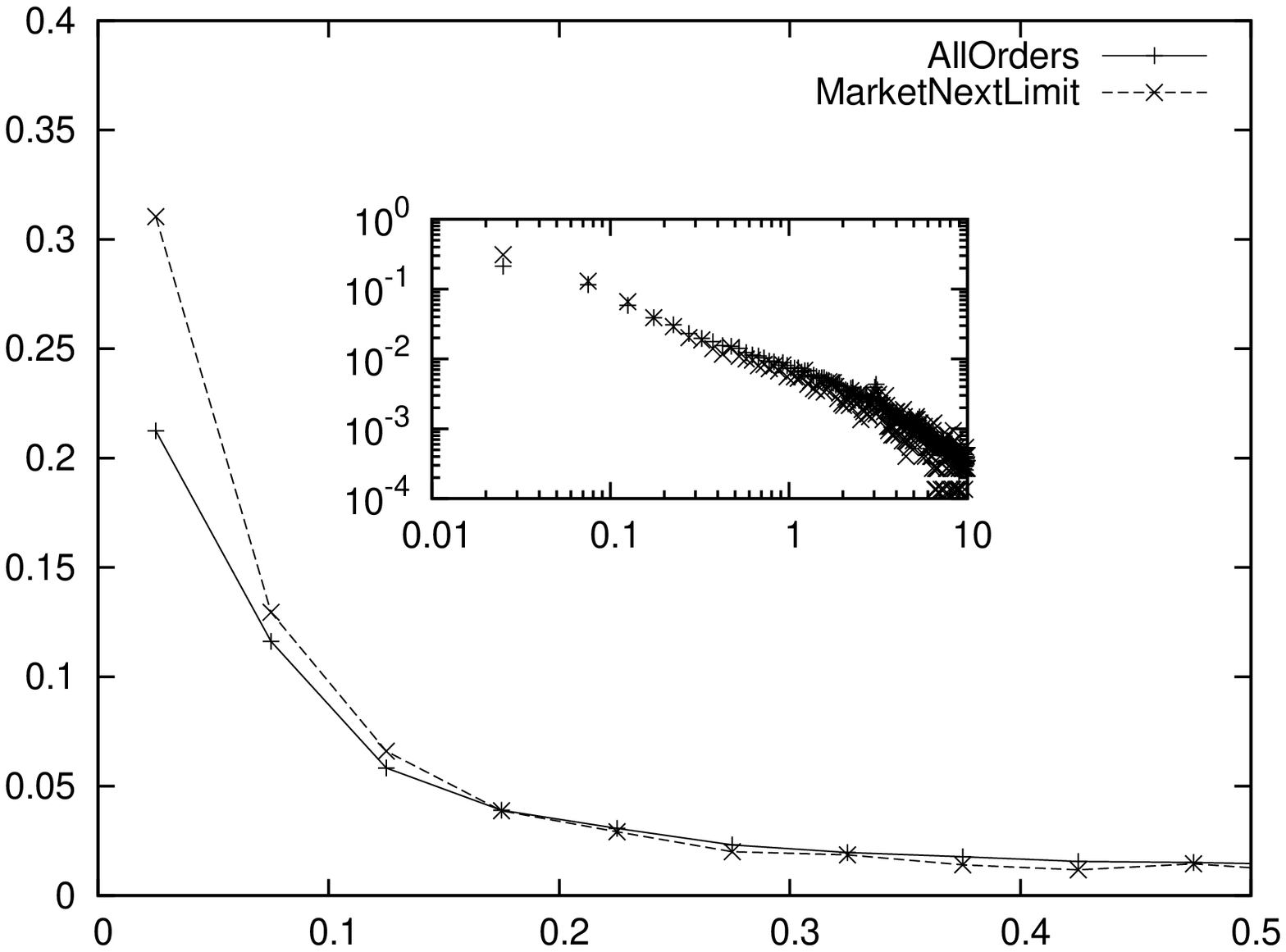}}
	&
	\hspace{-0.5cm}
	\rotatebox{270}{\includegraphics[width=0.365\textwidth]{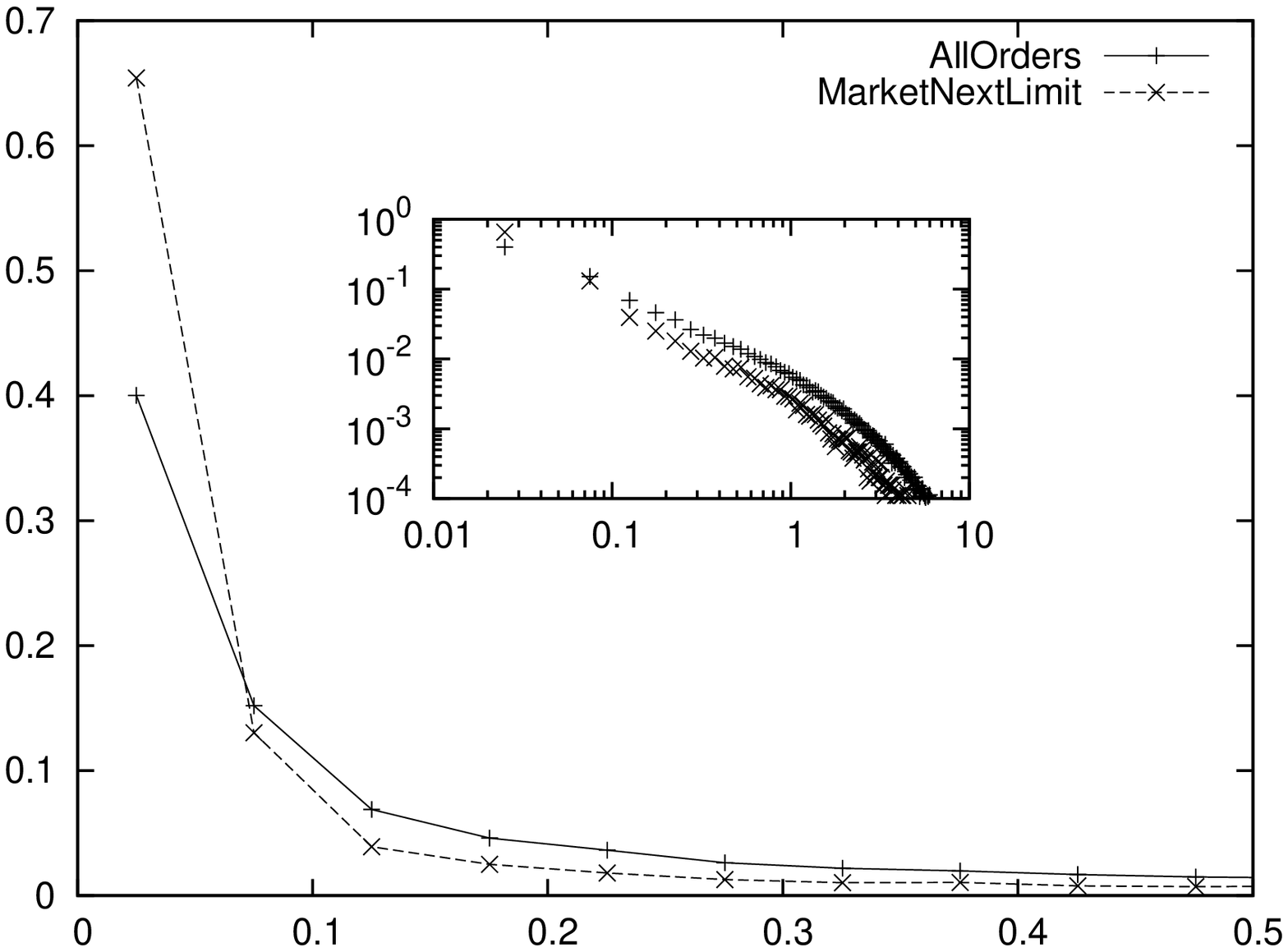}}	
\end{tabular}
\end{center}
\caption{Empirical density function of the distribution of the time intervals between two consecutive orders (any type, market or limit) and empirical density function of the distribution of the time intervals between a market order and the immediately following limit order. X-axis is scaled in seconds. In insets, same data using a log-log scale. Studied assets: BNPP.PA (top left), LAGA.PA (top right), FEIZ9 (bottom left), FFIZ9 (bottom right).}
\label{figure:EmpiricalPDF_AllOrdersInterArrivalTimes_MarketNextLimit}
\end{figure}

On the graphs shown in figure~\ref{figure:EmpiricalPDF_AllOrdersInterArrivalTimes_MarketNextLimit}, we plot the full empirical distributions for four of the five studied assets\footnote{Observations are identical on all the studied assets.}. We observe their broad distribution and the sharp peak for the shorter times: on the Footsie future market for example, 40\% of the measured time steps between consecutive events are less that 50 milliseconds ; this figure jumps to nearly 70\% when considering only market orders and their following limit orders.
This observation is an evidence for some sort of market-making behaviour of some participants on those markets. It appears that the submission of market orders is monitored and triggers automatic limit orders that add volumes in the order book (and not far from the best quotes, since we only monitor the five best limits).

In order to confirm this finding, we perform non-parametric statistical test on the measured data. For all four studied markets, omnibus Kolmogorov-Smirnov and Cramer-von Mises tests performed on random samples establish that the considered distributions are statistically different. If assuming a common shape, a Wilcoxon-Mann-Withney U test clearly states that the distribution of time intervals between a market orders and the following limit order is clearly shifted to the left compared to the distributions of time intervals between any orders, i.e. the average ``limit following market'' reaction time is shorter than the average time interval between random consecutive orders.

\begin{figure}[h!t]
\begin{center}
\begin{tabular}{cc}
	\hspace{-0.5cm}
	\rotatebox{270}{\includegraphics[width=0.365\textwidth]{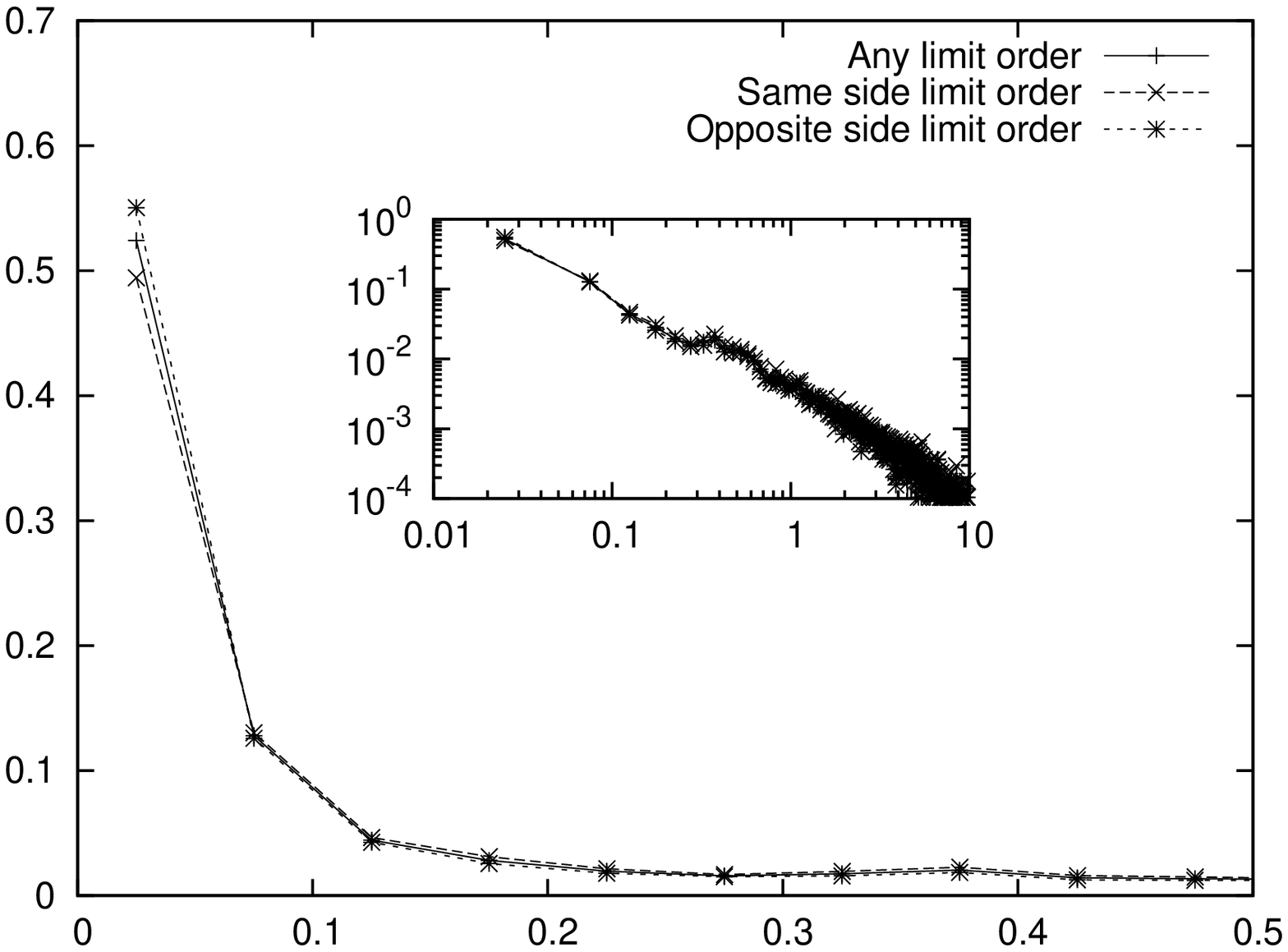}}
	&
	\hspace{-0.5cm}
	\rotatebox{270}{\includegraphics[width=0.365\textwidth]{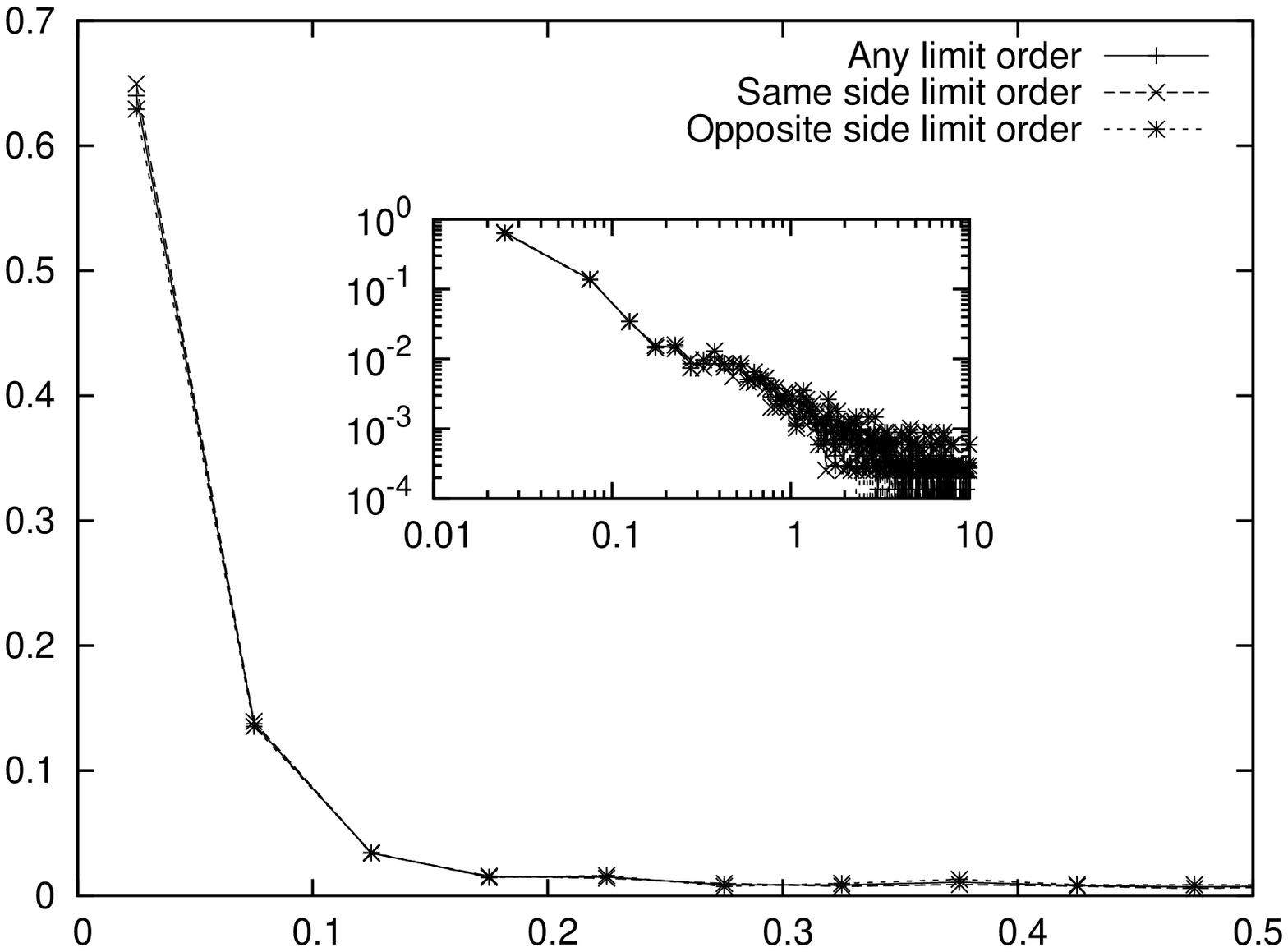}}
\end{tabular}
\end{center}
\caption{Empirical density function of the distribution of the time intervals between a market order and the immediately following limit order, whether orders have been submitted on the same side and on opposite sides. X-axis is scaled in seconds. In insets, same data using a log-log scale. Studied assets: BNPP.PA (left), LAGA.PA (right).}
\label{figure:EmpiricalPDF_SameOrOppositeSideLimitMarketInterArrivalTimes_MarketNextLimit}
\end{figure}

Note that there is no link with the sign of the market order and the sign of the following limit order. For example for the BNP Paribas (resp. Peugeot and Lagardere) stock, they have the same sign in 48.8\% (resp. 51.9\% and 50.7\%) of the observations.
And more interestingly, the ``limit following market'' property holds regardless of the side on which the following limit order is submitted. On figure~\ref{figure:EmpiricalPDF_SameOrOppositeSideLimitMarketInterArrivalTimes_MarketNextLimit}, we plot the empirical distributions of time intervals between a market order and the following limit order, conditionally on the side of the limit order: the same side as the market order or the opposite one. It appears for all studied assets that both distributions are roughly identical.
In other words, we cannot distinguish on the data if liquidity is added where the market order has been submitted or on the opposite side. Therefore, we do not infer any empirical property of placement: when a market order is submitted, the intensity of the limit order process increases \textit{on both sides} of the order book.

This effect we have thus identified is a phenomenon of liquidity replenishment of an order book after the execution of a trade. The fact that it is a bilateral effect makes its consequences similar to ``market making'', event though there is obviously no market maker involved on the studied markets.

\subsection{A reciprocal ``market following limit'' effect ?}
\label{subsection:EmpiricalReciprocal}

\begin{figure}[ht]
\begin{center}
\begin{tabular}{cc}
	\hspace{-0.5cm}
	\rotatebox{270}{\includegraphics[width=0.365\textwidth]{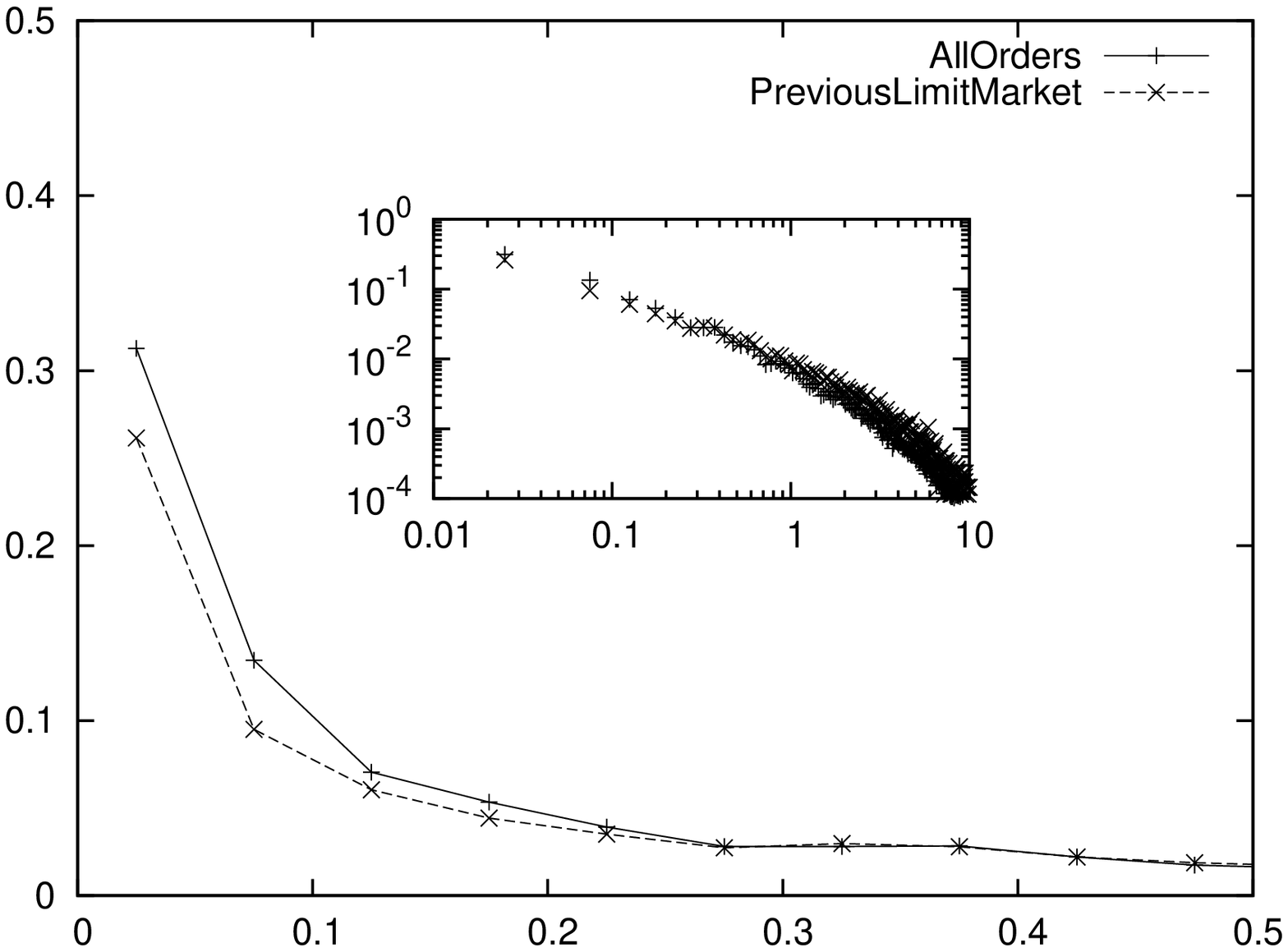}}
	&
	\hspace{-0.5cm}
	\rotatebox{270}{\includegraphics[width=0.365\textwidth]{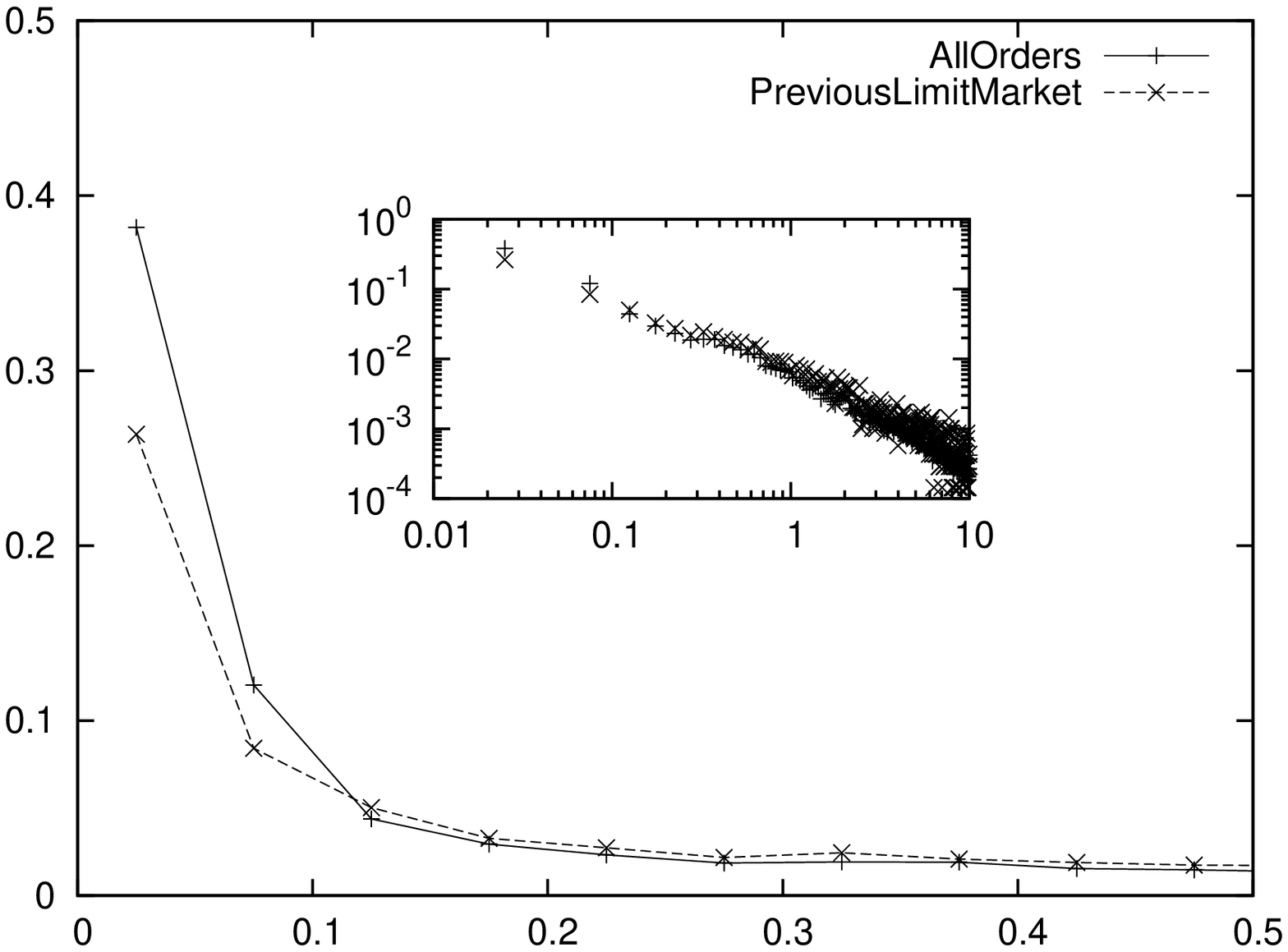}}
	\\
	\hspace{-0.5cm}
	\rotatebox{270}{\includegraphics[width=0.365\textwidth]{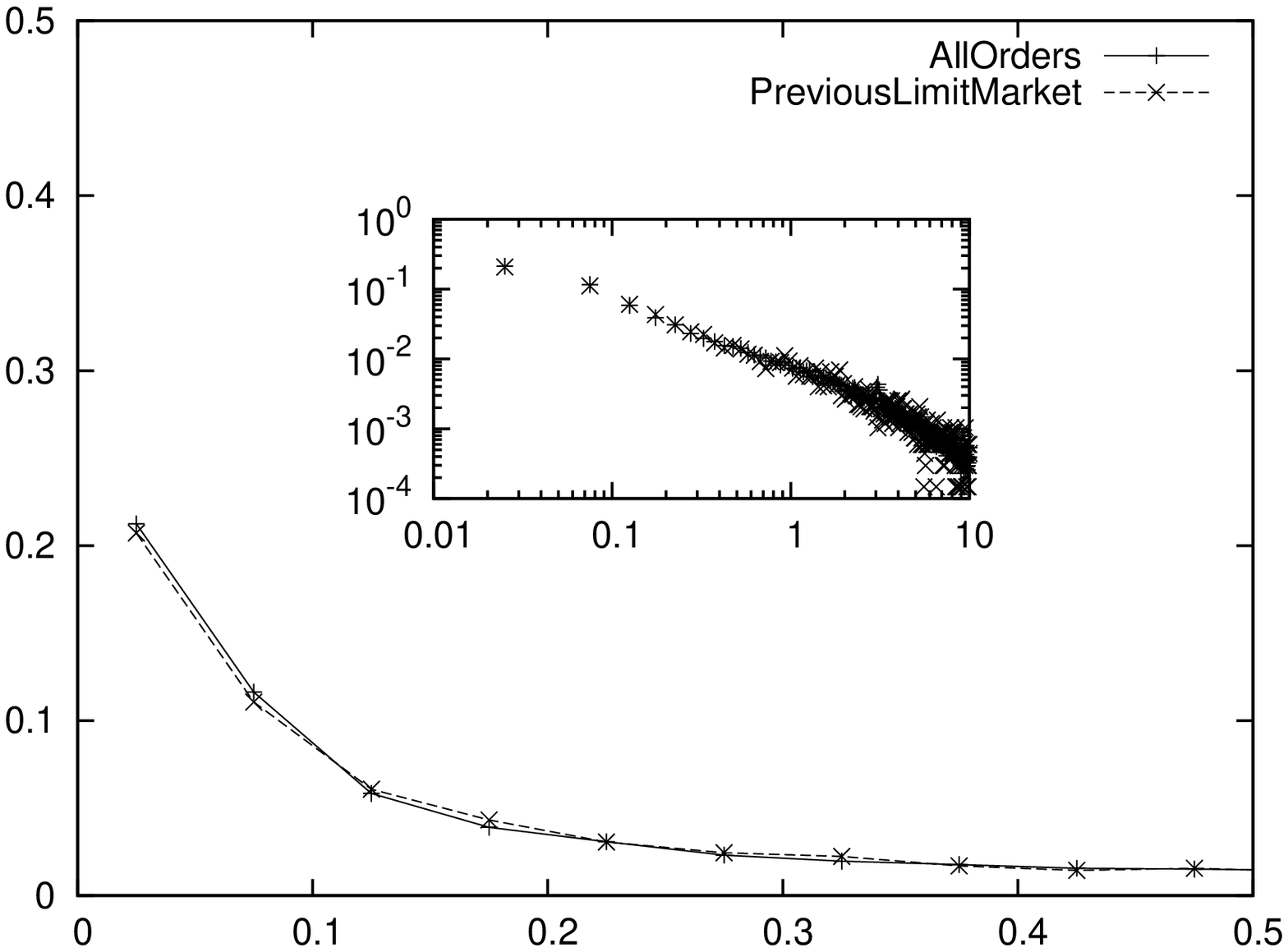}}
	&
	\hspace{-0.5cm}
	\rotatebox{270}{\includegraphics[width=0.365\textwidth]{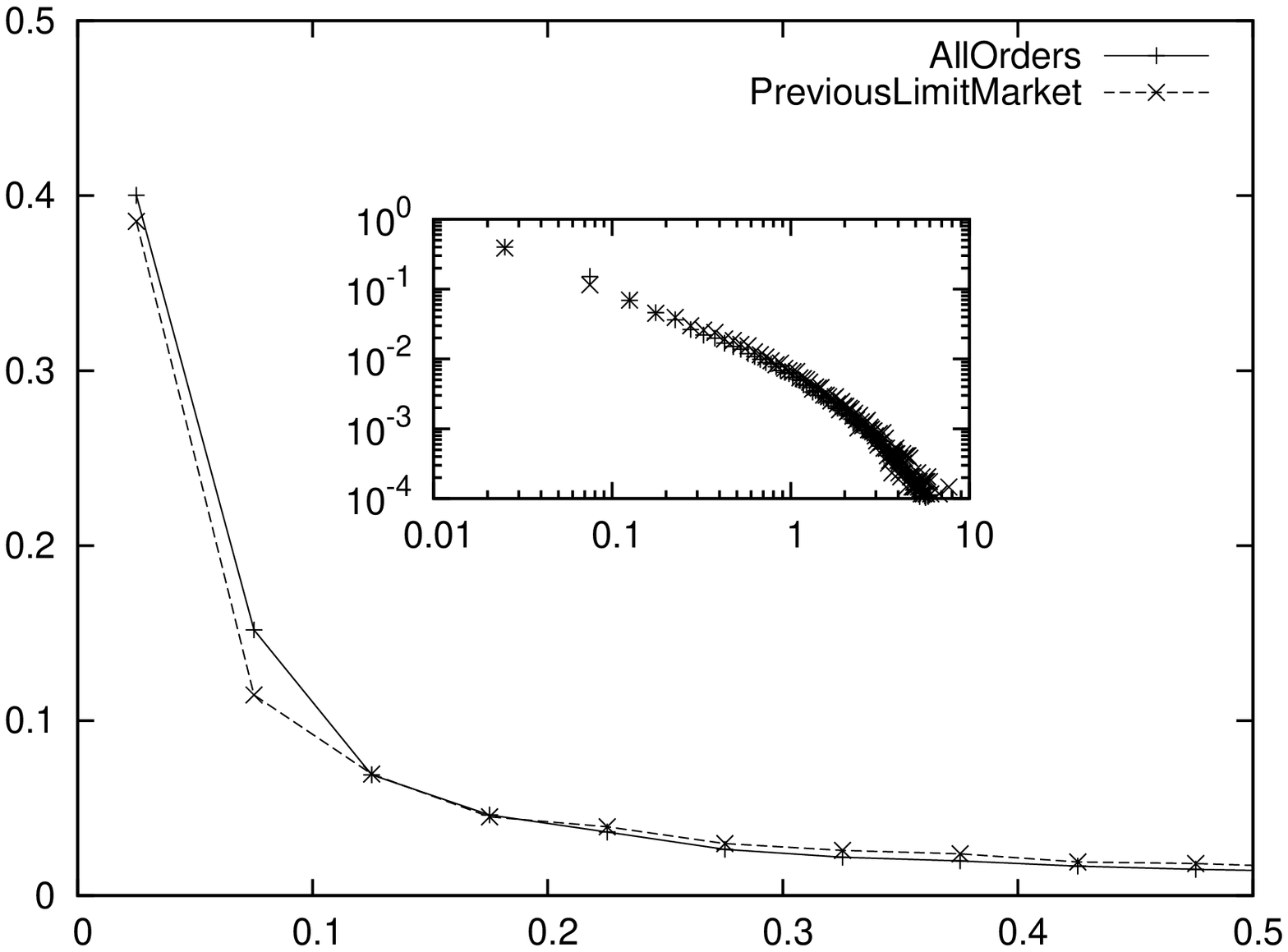}}	
\end{tabular}
\end{center}
\caption{Empirical density function of the distribution of the time intervals between two consecutive orders (any type, market or limit) and empirical density function of the distribution of the time intervals between a limit order and an immediately following market order. In insets, same data using a log-log scale. Studied assets: BNPP.PA (top left), LAGA.PA (top right), FEIZ9 (bottom left), FFIZ9 (bottom right).}
\label{figure:EmpiricalPDF_AllOrdersPreviousLimitMarketInterArrivalTimes}
\end{figure}

We now check if a similar or opposite distortion is to be found on market orders when they follow limit orders. To investigate this, we compute for all our studied assets the ``reciprocal'' measures:
\begin{itemize}
	\item the empirical probability density function (pdf) of the time intervals of the counting process of all orders (limit orders and market orders mixed), i.e. the time step between any order book event (other than cancellation)
	\item and the empirical density function of the time step between a market order and the previous limit order.
\end{itemize}
As previously, if an independent Poisson assumption held, then these empirical distribution should be identical. Results for four assets are shown on figure~\ref{figure:EmpiricalPDF_AllOrdersPreviousLimitMarketInterArrivalTimes}.
Contrary to previous case, no effect is very easily interpreted. For the three stocks (BNPP.PA, LAGA.PA and PEUP.PA (not shown)), it seems that the empirical distribution is less peaked for small time intervals, but difference is much less important than in the previous case. As for the FEI and FFI markets, the two distributions are even much closer. Non-parametric tests confirms these observations. 

Performed on data from the three equity markets, Kolmogorov tests indicate different distributions and Wilcoxon tests enforce the observation that time intervals between a limit order and a following market order are stochastically larger than time intervals between unidentified orders. As for the future markets on Footsie (FFI) and 3-month Euribor (FEI), Kolmogorov tests does not indicate differences in the two observed distributions, and the result is confirmed by a Wilcoxon test that concludes at the equality of the means.

\section{Order book models with mutually exciting order flows}
\label{section:OrderBookSimulator}
Following these previous observations, we enhance a basic agent-based order book simulator with dependence between the flows of limit and market orders.

\subsection{The basic Poisson model}
We use as base model a standard zero-intelligence agent-based market simulator built as follows. One agent is a liquidity provider. This agent submits limit orders in the order books, orders which he can cancel at any time. This agent is simply characterized by a few random distributions:
\begin{enumerate}
	\item submission times of new limit orders are distributed according to a homogeneous Poisson process $N^L$ with intensity $\lambda^L$;
	\item submission times of cancellation of orders are distributed according to homogeneous Poisson process $N^C$ with intensity $\lambda^C$;
	\item placement of new limit orders is centred around the same side best quote and follows a Student's distribution with degrees of freedom $\nu^P_1$, shift parameter $m^P_1$ and scale parameter $s^P_1$;
	\item new limit orders' volume is randomly distributed according to an exponential law with mean $m^V_1$;	
	\item in case of a cancellation, the agent deletes his own orders with probability $\delta$.
\end{enumerate}

The second agent in the basic model is a noise trader. This agent only submits market order (it is sometimes referred to as the liquidity taker). Characterization of this agent is even simpler:
\begin{enumerate}[resume]
	\item submission times of new market orders are distributed according to a homogeneous Poisson process $N^M$ with intensity $\mu$;
	\item market orders' volume is randomly distributed according to an exponential law with mean $m^V_2$.
\end{enumerate}

For all the experiments, agents submit orders on the bid or the ask side with probability 0.5. This basic model will be henceforth referred to as ``HP'' (Homogeneous Poisson).

Assumptions 1, 2 and 6 (Poisson) will be replaced in our enhanced model. Assumption 3 (Student) is in line with empirical observations in \cite{MikeFarmer2008}. Assumptions 4 and 7 are in line with empirical observations in \cite{ChakrabortiMuniToke2009} as far as the main body of the distribution is concerned, but fail to represent the broad distribution observed in empirical studies. All the parameters except $\delta$, which we kept exogenous, can be more or less roughly estimated on our data. In fact $\delta$ is the parameter of the less realistic feature of this simple model, and is thus difficult to calibrate. It can be used as a free parameter to fit the realized volatility.

\subsection{Adding dependence between order flows}
We have found in section~\ref{subsection:EmpiricalEvidence} that market data shows that the flow of limit orders strongly depends on the flow of market order. We thus propose that in our experiment, the flow of limit and market orders are modelled by Hawkes processes $N^L$ and $N^M$, with stochastic intensities respectively $\lambda$ and $\mu$ defined as:
\begin{equation}
\displaystyle
\left\{
\begin{array}{rcccc}
	\displaystyle\mu^M(t) & = & \displaystyle \mu^M_0 & + \displaystyle\int_0^t \alpha_{MM}e^{-\beta_{MM}(t-s)} dN^M_s &
	\\
	\displaystyle\lambda^L(t) & = & \displaystyle \lambda^L_0 & + \displaystyle\int_0^t \alpha_{LM}e^{-\beta_{LM}(t-s)} dN^M_s & + \displaystyle\int_0^t \alpha_{LL}e^{-\beta_{LL}(t-s)} dN^L_s
\end{array}
\right.
\end{equation}

Three mechanisms can be used here. The first two are self-exciting ones, ``MM'' and ``LL''. They are a way to translate into the model the observed clustering of arrival of market and limit orders and the broad distributions of their durations. In the empirical study \cite{Large2007}, it is found that the measured excitation MM is important. In our simulated model, we will show (see \ref{subsection:Results_ArrivalTimes}) that this allows a simulator to provide realistic distributions of the durations of trades.

The third mechanism, ``LM'', is the direct translation of the empirical property we have presented in section~\ref{subsection:EmpiricalEvidence}. When a market order is submitted, the intensity of the limit order process $N^L$ increases, enforcing the probability that a ``market making'' behaviour will be the next event. We do no implement the reciprocal mutual excitation ``ML'', since we do not observe that kind of influence on our data, as explained in section \ref{subsection:EmpiricalReciprocal}.

Rest of the model is unchanged. Turning these features successively on and off gives us several models to test -- namely HP (Homogeneous Poisson processes), LM, MM, MM+LM, MM+LL, MM+LL+LM -- to try to understand the influence  of each effect.

\section{Numerical results on the order book}
\label{section:NumericalResults}

\subsection{Fitting and simulation}
\label{subsection:Fitting}
We fit these processes by computing the maximum likelihood estimators of the parameters of the different models on our data. As expected, estimated values varies with the market activity on the day of the sample. However, it appears that estimation of the parameters of stochastic intensity for the MM and LM effect are quite robust. We find an average relaxation parameter $\hat\beta_{MM}=6$, i.e. roughly $170$ milliseconds as a characteristic time for the MM effect, and $\hat\beta_{LM}=1.8$, i.e. roughly $550$ milliseconds characteristic time for the LM effect. Estimation of models including the LL effect are much more troublesome on our data. In the simulations that follows, we assume that the self-exciting parameters are similar ($\alpha_{MM}=\alpha_{LL}$, $\beta_{MM}=\beta_{LL}$) and ensure that the number of market orders and limit orders in the different simulations is roughly equivalent (i.e. approximately 145000 limit orders and 19000 market orders for 24 hours of continuous trading). Table~\ref{table:ParametersPerModel} summarizes the numerical values used for simulation. Fitted parameters are in agreement with an assumption of asymptotic stationarity.
\begin{table}[ht]
\begin{center}
\begin{tabular}{|c|c|cc|c|cc|cc|}
	\hline
	Model & $\mu_0$ & $\alpha_{MM}$ & $\beta_{MM}$ & $\lambda_0$ & $\alpha_{LM}$ & $\beta_{LM}$ & $\alpha_{LL}$ & $\beta_{LL}$ \\
	\hline
	HP & 0.22 & - & - & 1.69 & - & - & - & - \\
	\hline
	LM & 0.22 & - & - & 0.79 & 5.8 & 1.8 & - & - \\
	\hline
	MM & 0.09 & 1.7 & 6.0 & 1.69 & - & - & - & - \\
	\hline
	MM LL & 0.09 & 1.7 & 6.0 & 0.60 & - & - & 1.7 & 6.0 \\
	\hline
	MM LM & 0.12 & 1.7 & 6.0 & 0.82 & 5.8 & 1.8 & - & - \\
	\hline
	MM LL LM & 0.12 & 1.7 & 5.8 & 0.02 & 5.8 & 1.8 & 1.7 & 6.0 \\
	\hline	
\end{tabular}
\begin{tabular}{|rl|}
	\hline
	Common parameters: & $m^P_1=2.7, \nu^P_1=2.0, s^P_1=0.9$ \\
	& $^V_1=275, m^V_2=380$ \\
	& $\lambda^C=1.35, \delta = 0.015 $ \\
	\hline
\end{tabular}
\end{center}
\caption{Estimated values of parameters used for simulations.}
\label{table:ParametersPerModel}
\end{table}

We compute long runs of simulations with our enhanced model, simulating each time 24 hours of continuous trading. Note that using the chosen parameters, we never face the case of an empty order book. We observe several statistics on the results, which we discuss in the following sections.

\subsection{Impact on arrival times}
\label{subsection:Results_ArrivalTimes}
\begin{figure}[h!t]
\begin{center}
\begin{tabular}{cc}
	\hspace{-0.5cm}
	\rotatebox{270}{\includegraphics[width=0.365\textwidth]{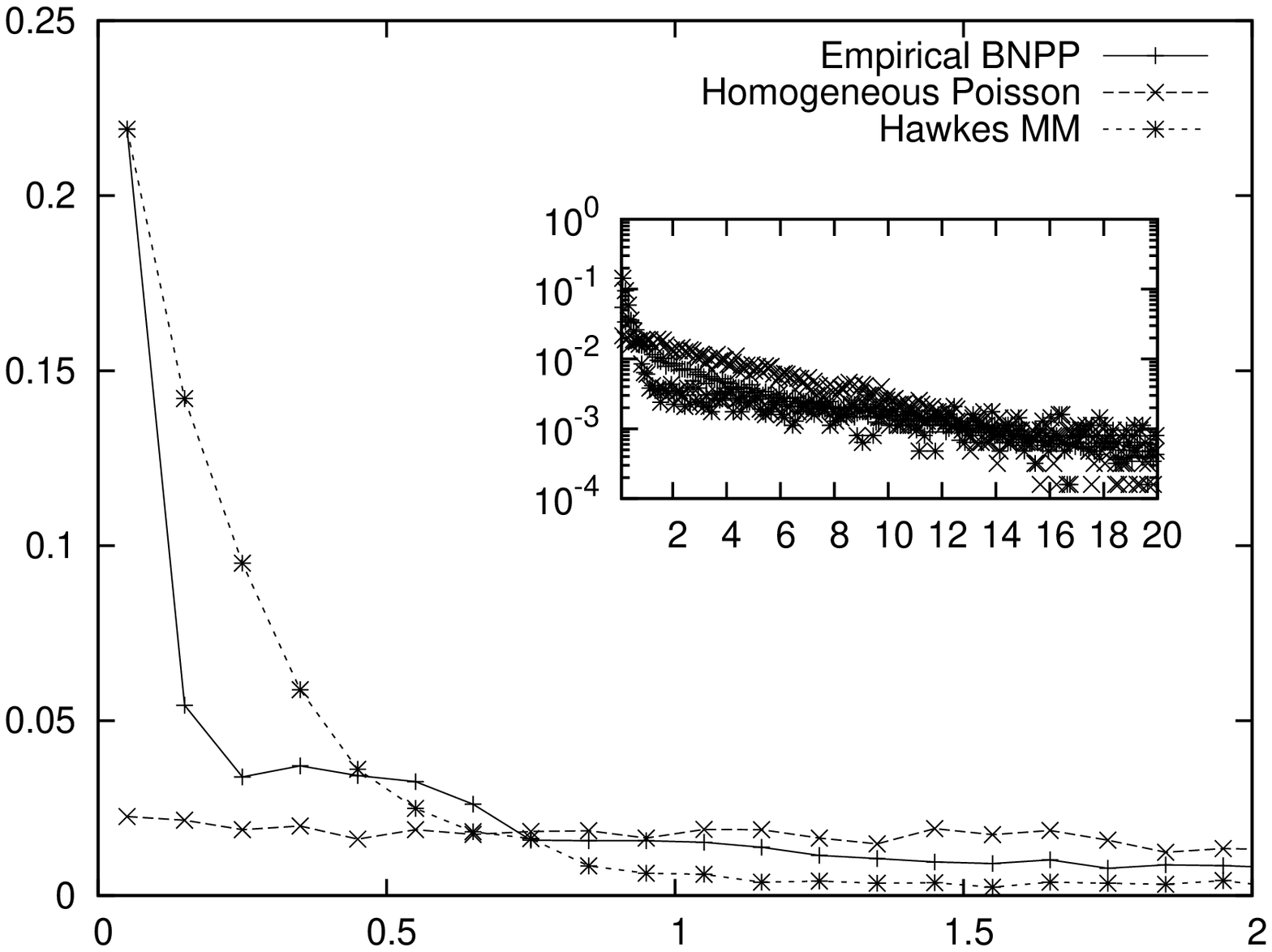}}
	&
	\hspace{-0.5cm}
	\rotatebox{270}{\includegraphics[width=0.365\textwidth]{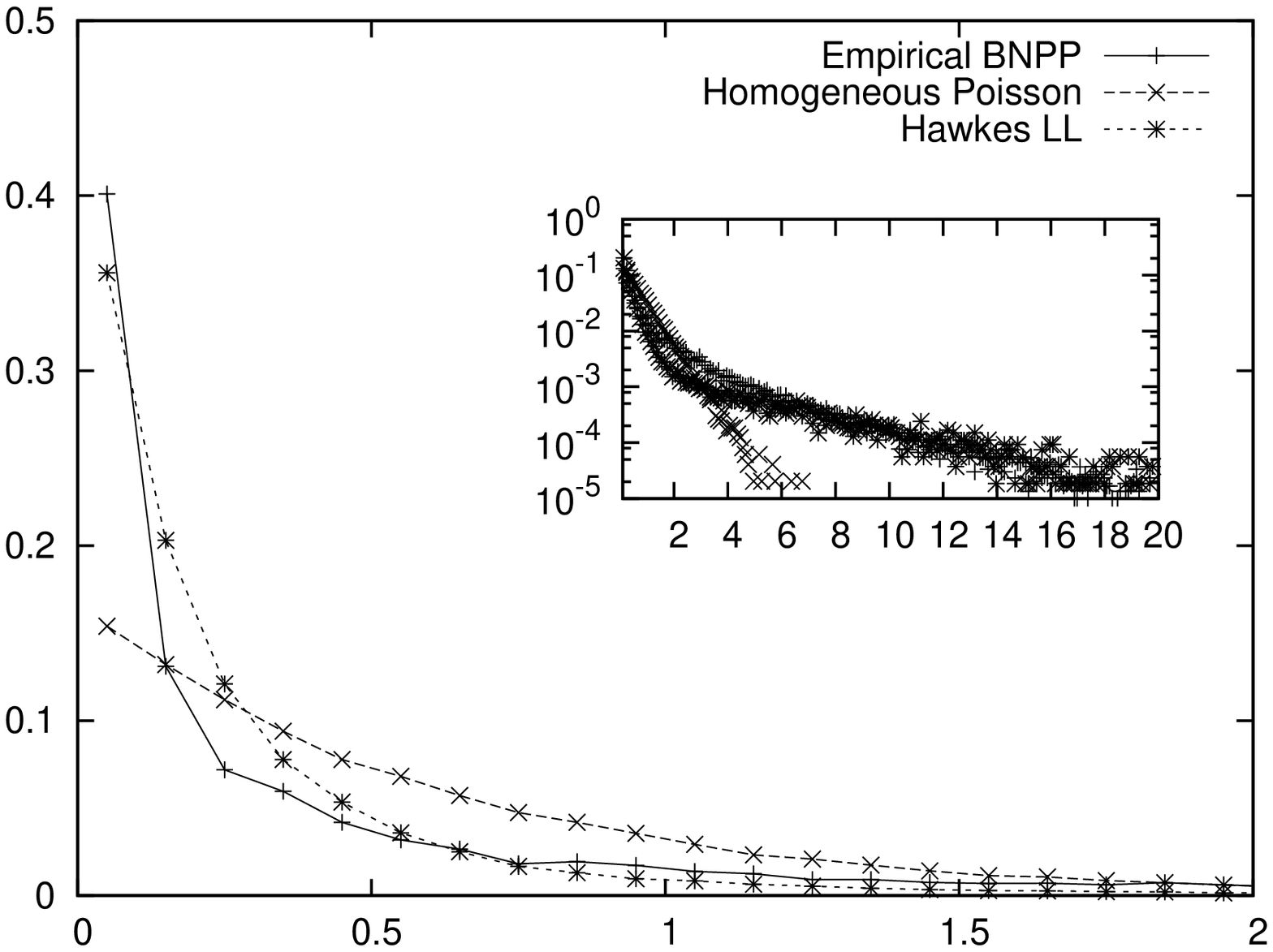}}
\end{tabular}
\end{center}
\caption{Empirical density function of the distribution of the durations of market orders (left) and limit orders (right) for three simulations, namely HP, MM, LL, compared to empirical measures. In inset, same data using a semi-log scale. }
\label{figure:ArrivalTimes_Emp_HP_MM_LL}
\end{figure}

We can easily check that introducing self- and mutually exciting processes into the order book simulator helps producing more realistic arrival times. Figure~\ref{figure:ArrivalTimes_Emp_HP_MM_LL} shows the distributions of the durations of market orders (left) and limit orders (right). As expected, we check that the Poisson assumption has to be discarded, while the Hawkes processes help getting more weight for very short time intervals.

\begin{figure}[h!t]
\begin{center}
\rotatebox{270}{\includegraphics[width=0.45\textwidth]{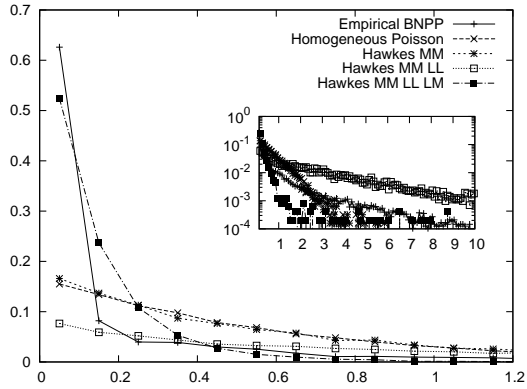}}
\end{center}
\caption{Empirical density function of the distribution of the time intervals between a market order and the following limit order for three simulations, namely HP, MM+LL, MM+LL+LM, compared to empirical measures. In inset, same data using a semi-log scale. }
\label{figure:MarketNextLimitArrivalTimes_Emp_HP_LL_MMLL_MMLLLM}
\end{figure}

We also verify that models with only self-exciting processes MM and LL are not able to reproduce the ``liquidity replenishment'' feature described in section~\ref{subsection:EmpiricalEvidence}. Distribution of time intervals between a market order and the next limit order are plotted on figure \ref{figure:MarketNextLimitArrivalTimes_Emp_HP_LL_MMLL_MMLLLM}. As expected, no peak for short times is observed if the LM effect is not in the model. But when the LM effect is included, the simulated distribution of time intervals between a market order and the following limit order is very close to the empirical one.

\subsection{Impact on the bid-ask spread}
\label{subsection:Results_Spread}
\begin{figure}[h!t]
\begin{center}
\rotatebox{270}{
\includegraphics[width=0.5\textwidth]{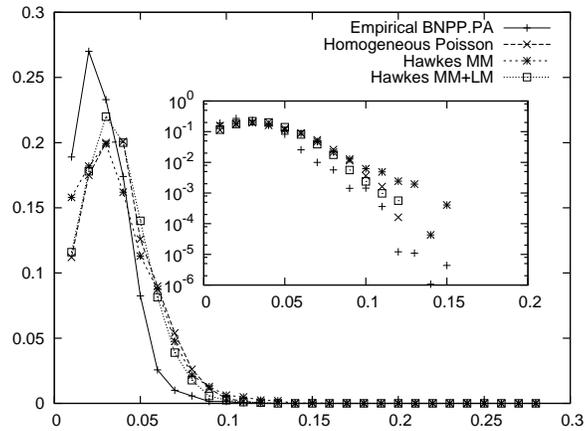}	
}
\end{center}
\caption{Empirical density function of the distribution of the bid-ask spread for three simulations, namely HP, MM, MM+LM, compared to empirical measures. In inset, same data using a semi-log scale. X-axis is scaled in euro (1 tick is 0.01 euro). }
\label{figure:Spread_Emp_HP_MM_MMLM_BNPP}
\end{figure}

Besides a better simulation of the arrival times of orders, we argue that the LM effect also helps simulating a more realistic behaviour of the bid-ask spread of the order book. On figure~\ref{figure:Spread_Emp_HP_MM_MMLM_BNPP}, we compare the distributions of the spread for three models -- HP, MM, MM+LM -- in regard to the empirical measures.
We first observe that the model with homogeneous Poisson processes produces a fairly good shape for the spread distribution, but slightly shifted to the right. Small spread values are largely underestimated. When adding the MM effect in order to get a better grasp at market orders' arrival times, it appears that we flatten the spread distribution. One interpretation could be that when the process $N^M$ is excited, markets orders tend to arrive in cluster and to hit the first limits of the order book, widening the spread and thus giving more weight to large spread values. But since the number of orders is roughly constant in our simulations, there has to be periods of lesser market activity where limit orders reduce the spread. Hence a flatter distribution.

Here, the MM+LM model produces a spread distribution much closer to the empirical shape. It appears from figure~\ref{figure:Spread_Emp_HP_MM_MMLM_BNPP} that the LM effect reduces the spread: the ``market making'' behaviour, i.e. limit orders triggered by market orders, helps giving less weight to larger spread values (see the tail of the distribution)  and to sharpen the peak of the distribution for small spread values. Thus, it seems that simulations confirm the empirical properties of a ``market making'' behaviour on electronic order books.

\begin{figure}[h!t]
\begin{center}
\rotatebox{270}{
\includegraphics[width=0.5\textwidth]{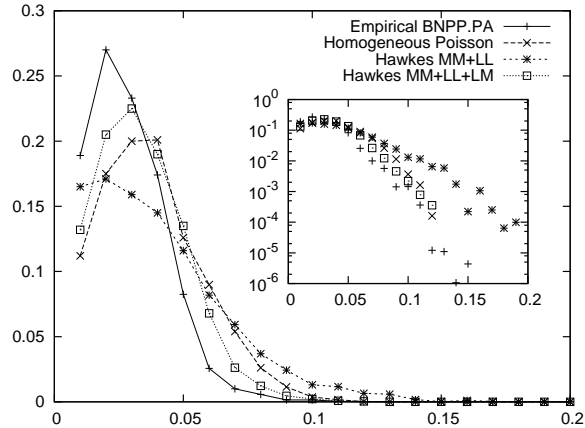}	
}
\end{center}
\caption{Empirical density function of the distribution of the bid-ask spread three simulations, namely HP, MM, MM+LM, compared to empirical measures. In inset, same data using a semi-log scale. X-axis is scaled in euro (1 tick is 0.01 euro). }
\label{figure:Spread_Emp_HP_MMLL_MMLLLM_BNPP}
\end{figure}

We show on figure~\ref{figure:Spread_Emp_HP_MMLL_MMLLLM_BNPP} that the same effect is observed in an even clearer way with the MM+LL and MM+LL+LM models. 
Actually, the spread distribution produced by the MM+LL model is the flattest one. This is in line with our previous argument. When using two independent self exciting Hawkes processes for arrival of orders, periods of high market orders' intensity gives more weight to large spread values, while periods of high limit orders' intensity gives more weight to small spread values. Adding the cross-term LM to the processes implements a coupling effect that helps reproducing the empirical shape of the spread distribution. The MM+LL+LM simulated spread is the closest to the empirical one.

\subsection{A remark on price returns in the model}
\begin{figure}[ht]
\begin{center}
\rotatebox{270}{
\includegraphics[width=0.5\textwidth]{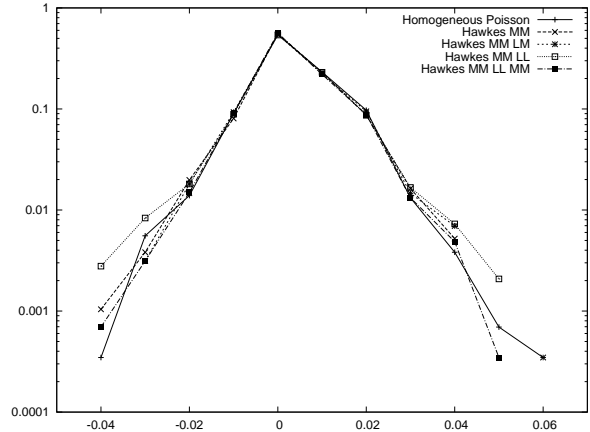}	
}
\end{center}
\caption{Empirical density function of the distribution of the 30-second variations of the mid-price for five simulations, namely HP, MM, MM+LM, MM+LL, MM+LL+LM, using a semi-log scale. X-axis is scaled in euro (1 tick is 0.01 euro). }
\label{figure:EmpiricalPDF_SimulatedDeltaMidPrice30}
\end{figure}

It is somewhat remarkable to observe that these variations of the spread distributions are obtained with little or no change in the distributions of the variations of the mid-price. As shown on figure~\ref{figure:EmpiricalPDF_SimulatedDeltaMidPrice30}, the distributions of the variations of the mid-price sampled every 30 seconds are nearly identical for all the simulated models (and much tighter than the empirical one).
This is due to the fact that the simulated order books are much more furnished than the empirical one, hence the smaller standard deviation of the mid price variations. One solution to get thinner order books and hence more realistic values of the variations of the mid-price would be to increase our exogenous parameter $\delta$. But in that case, mechanisms for the replenishment of an empty order book should be carefully studied, which is still to be done.

\section{Conclusion}
We have shown the the use of Hawkes processes may help producing a realistic shape of the spread distribution in an agent-based order book simulator. We emphasize on the role of the excitation of the limit order process by the market order process. This coupling of the processes, similar to a ``market making'' behaviour, is empirically observed on several markets, and simulations confirms it is a key component for realistic order book models.

Future work should investigate if other processes or other kernels ($\nu_{LM}$ in our notation) might better fit the observed orders flows. In particular, we observe very short characteristic times, which should lead us to question the use of the exponential decay. Furthermore, as pointed out in the paper, many other mechanisms are to be investigated: excitation of markets orders, link with volumes, replenishment of an empty order book, etc.

\bibliographystyle{amsplain}
\bibliography{epkolv}

\end{document}